# Electromechanical Switching and Momentum-Selective Transport in Geometry-Defined Blue Phosphorus Homojunctions

Zewen Wu,[1] Min Zhou,[2] Yanxia Xing,[2,a)] Xianghua Kong[1,a)]

[1] *College of Physics and Optoelectronic Engineering, Shenzhen University, Shenzhen 518060, China*
[2] *School of Physics, Beijing Institute of technology, Beijing, 100081, China*
(*Electronic mail: xingyanxia@bit.edu.cn; kongxianghuaphysics@szu.edu.cn.)

**ABSTRACT**

Developing intrinsic homojunctions without chemical heterogeneity remains a key challenge in future two-dimensional devices. Here, we report a geometry-defined metal–semiconductor–metal homojunction in bilayer blue phosphorus (BlueP) created by a localized bubble corrugation, without chemical doping or foreign-material interfaces. First-principles calculations show that enlarging the interlayer separation in the metallic $A_1B_{-1}$-stacked BlueP bilayer opens a band gap, enabling a semiconducting barrier embedded between metallic segments. First-principles quantum-transport simulations reveal a crossover from ballistic to tunneling transport upon bubble formation. In the tunneling regime, transmission decreases exponentially with bubble width while remaining weakly sensitive to bubble height and bulging direction. The junction acts as an orientation-dependent $k$-space filter, producing transport anisotropy and momentum selectivity. Orbital-resolved scattering analysis shows that intralayer-bonding channels persist under deformation whereas interlayer-hybridized channels are quenched, and that $\sigma$-type bonding yields higher conductance than $\pi$-type bonding. These insights motivate two electromechanical device concepts: a mechanically switchable memory element with ON/OFF ratios up to 30 and a nanoscale sliding rheostat with reproducible exponential resistance tuning for Ångström-scale displacement sensing.

Two-dimensional (2D) materials are promising building blocks for next-generation electronic and optoelectronic devices.[1-6] Their atomically thin nature enables strong electrostatic gate control and helps suppress short-channel effects.[7-10] A wide range of proof-of-concept devices has been demonstrated, including light-emitting diodes,[11-13] photodetectors,[14-16] and field-effect transistors.[1,2,4-9,17-20] However, achieving high performance in practical 2D devices remains challenging, and a major limitation lies at the interface between the 2D channel and electrodes or surrounding materials. In typical device architectures, the 2D semiconductor is contacted by bulk metals or integrated with dissimilar 2D layers, forming heterogeneous interfaces. These interfaces can host interfacial states and induce Fermi-level pinning and local structural/electrostatic inhomogeneity, leading to Schottky barriers, contact resistance, and carrier scattering.[21-26] These parasitic effects become increasingly detrimental as device dimensions are scaled down.

Homojunctions offer a potential route to mitigate interface-related limitations through defining the conductive electrode regions and the semiconducting channel within a single, continuous material. By preserving lattice continuity and avoiding chemical mismatch, homojunctions can reduce extrinsic scattering and enable carrier separation without introducing a foreign-material junction interface. Currently, creating such junctions relies primarily on chemical doping, electrostatic split-gating, or phase engineering.[27-32] While effective, chemical doping can introduce disorder and spatial inhomogeneity and can be difficult to reversibly control at the nanoscale; electrostatic gating requires patterned local-gate architectures that increase device complexity and hinder high-density integration; and phase engineering can involve structural reconstruction at phase boundaries, which may introduce additional scattering.[27,29,32] Therefore, developing a chemistry-free and materials-intrinsic strategy to define high-quality homojunctions while maintaining lattice continuity remains highly desirable.

For 2D materials, electronic structures can be significantly modulated by the layer number and stacking registry, because interlayer hybridization provides an efficient handle to reshape band dispersions and the electronic phase. Blue phosphorus (BlueP) is an appealing system in this regard. Owing to its buckled honeycomb lattice and strong interlayer interaction, BlueP is predicted to transition from a semiconducting monolayer to a metallic bilayer in the energetically favorable bilayer $A_1B_{-1}$ stacking.[33-35] This thickness- and stacking-dependent contrast naturally motivates the possibility of using geometry to define homojunctions: if one can locally modulate the interlayer separation and/or stacking registry through controlled corrugations (e.g., nanoscale bubbles), then metallic and semiconducting regions may be patterned laterally within a single BlueP bilayer, enabling all-BlueP junctions with electrode and channel segments defined by geometry structure rather than chemistry. Beyond this intrinsic tunability, BlueP has been experimentally synthesized and has been reported to possess device-relevant electronic characteristics,[36-40] motivating a systematic investigation of charge transport in geometry-defined BlueP homojunctions, whose transport physics and device potential remain unexplored.

In this work, we propose a geometry-defined homojunction scheme in bilayer BlueP, where local bending and layer separation modulate interlayer coupling and intralayer bonding. Density functional theory (DFT) show that, in the $A_1B_{-1}$ stacking BlueP bilayer, increasing the interlayer separation weakens the interlayer coupling, opens a band gap, and drives a metal-to-semiconductor transition. Guided by this mechanism, we introduce a localized bubble-like deformation that creates a semiconducting barrier between metallic bilayer segments, thereby forming an all-BlueP metal-semiconductor-metal homojunction without chemical doping or foreign-material interfaces. Charge transport is then evaluated using first-principles quantum-transport simulations within the Non-Equilibrium Green's Function–Density Functional Theory (NEGF-DFT) framework. We find a crossover from ballistic to tunneling-dominated transport. In the tunneling regime, the transmission decays exponentially with the effective bubble width but depends weakly on bubble height or bulging direction. The homojunctions further exhibit orientation-dependent anisotropy and momentum-selective filtering in k-space. Orbital-resolved analysis attributes these trends to the suppression of interlayer-hybridized channels, the resilience of intralayer-bonding channels, and the higher conductance supported by $\sigma$-type bonding relative to $\pi$-type bonding. Finally, we outline two electromechanical device concepts enabled by geometry-controlled transport: a mechanically switchable memory element with a large ON/OFF ratio and a nanoscale sliding rheostat enabling reproducible resistance tuning and Ångström-scale displacement sensing, providing practical design guidelines for geometry-defined homojunction nanodevices in 2D bilayers.

The electronic properties of bilayer BlueP are critically dependent on its stacking configuration and interlayer interaction. As shown in Fig. S1(a), the energy-optimized $A_1B_{-1}$ stacking features an interlayer distance ($d$) of 3.01 Å. This configuration exhibits a metallic character, in distinct contrast to the semiconducting monolayer which possesses a band gap of 0.98 eV. Crucially, our calculations reveal that increasing the interlayer distance effectively decouples the layers, inducing a metal-to-semiconductor transition at a critical distance of $d$ = 4.51 Å (Fig. S2). This tunable dependence of the electronic phase on layer separation provides the fundamental physical basis for constructing homojunctions by locally separating the layers.

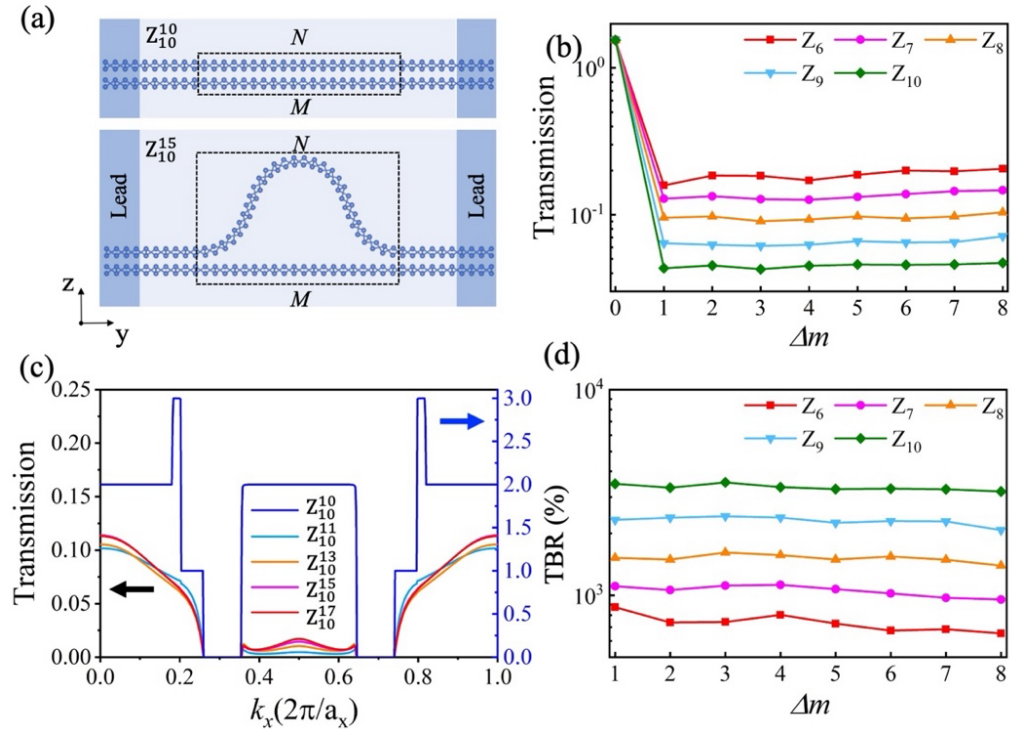

**FIG. 1.** Atomic structures and performances of zigzag homojunctions. (a) Atomic structures of two typical homojunctions marked with structural parameters (N, M is the number of unit cells contained in each layer within the dashed boxes). (b) Transmission versus $\Delta m$ ($\Delta m = N - M$); (c) Transmission versus $k_x$ for $Z_{10}^N (N \in [10,17])$, the blue line corresponds to the right Y-axis, the others to the left (arrows). (d) Tunneling Bend Resistance ratio.

Based on the interlayer-separation-driven metal–semiconductor transition, we construct lateral homojunctions by introducing a localized bubble-like deformation in the channel region of BlueP bilayers. As illustrated in Figs. 1(a) and S3(a), the device consists of a planar bottom layer and a locally arched top layer. The resulting local layer separation creates a semiconducting scattering region embedded between metallic bilayer segments, which serve as electrodes, forming an all-blue-phosphorus metal–semiconductor–metal homojunction. Charge transport is investigated along both the zigzag (Z) and armchair (A) crystallographic orientations. To quantify the bubble geometry, we define $M$ as the number of unit cells spanned by the bubble in the planar bottom layer, which determines the bubble width along the transport direction. $N$ represents the number of unit cells along the arched top layer within the same region, where the difference $\Delta m = N - M$ governs the curvature and effective height of the bubble. For brevity, each configuration is labeled as $X_M^N$, where $X$ indicates the transport orientation ($Z$ or $A$). For example, in the lower panel of Fig. 1(a), $Z_{10}^{15}$ corresponds to a zigzag-oriented junction with a bubble width of 10 unit cells ($M = 10$) and an arch length of 15 unit cells ($N = 15$).

Transport calculations were carried out for a series of bubble-defined junctions $X_M^N$ by varying the bubble width parameter $M$ from 6 to 10 and the curvature parameter $\Delta m$ from 0 to 8 for both zigzag and armchair transport directions. For each configuration, we evaluate the transmission coefficient within the NEGF-DFT framework. To quantify the ON/OFF contrast between the bubble-free metallic junction $T_M^M$ and the bubble-containing junction $T_M^N$, we introduce the tunneling bend resistance ratio (TBR), defined as:

$$TBR = \frac{T_M^M - T_M^N}{T_M^N} \times 100\%$$

where $T_M^M$ and $T_M^N$ are the transmission of $X_M^M$ and $X_M^N$, respectively.

Transport results reveal two distinct regimes controlled by bubble formation. We first discuss the zigzag orientation to establish the mechanism. In the absence of a bubble ($\Delta m = 0$), all $Z_M^M$ devices exhibit nearly identical transmission ($T = 1.58$), and the step-like spectrum in Fig. 1(c) is characteristic of ballistic conduction. Introducing a minimal bubble ($\Delta m : 0 \to 1$) suppresses the transmission by more than an order of magnitude, signaling the onset of tunneling transport. This abrupt reduction originates from the local metal-to-semiconductor conversion in the bubble region: the interlayer separation there increases to $\geq 6.32$Å, well above the critical distance of 4.51 Å required for gap opening.

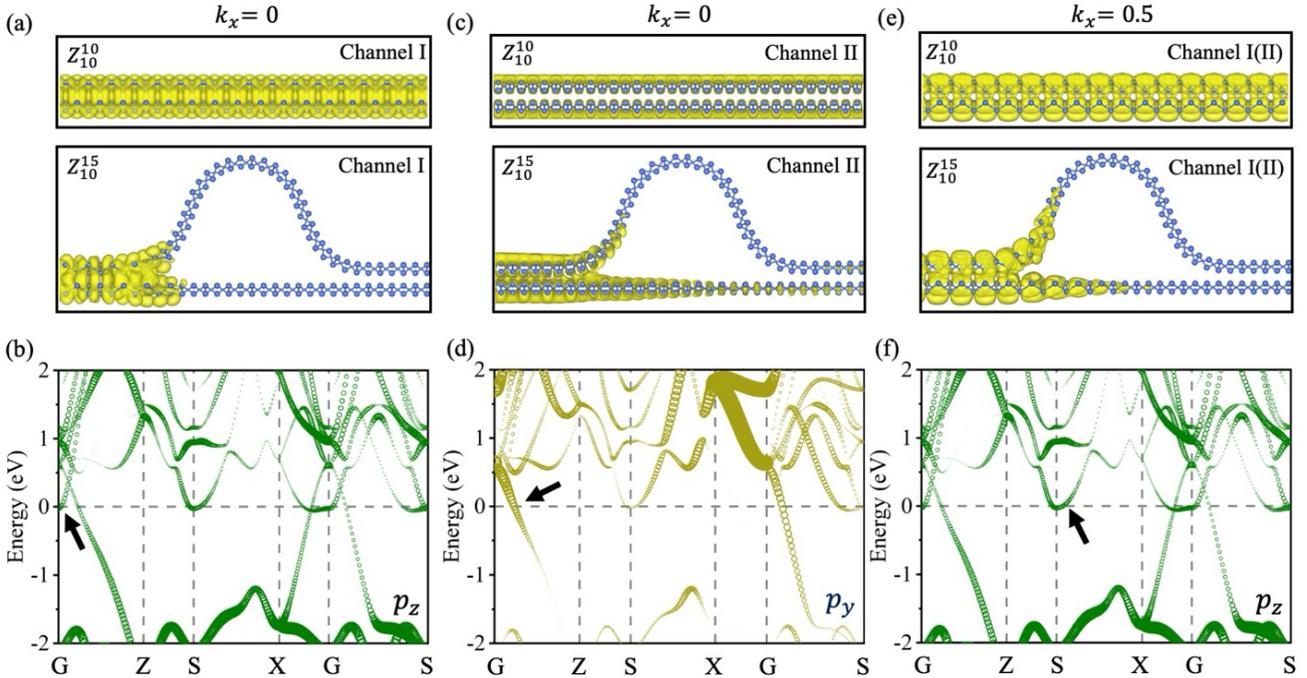

**FIG. 2.** Originate of momentum-selective filtering effect in zigzag homojunctions. (a, c, e) Scattering states for transistor $Z_{10}^{10}$ (upper panel) and $Z_{10}^{15}$ (lower panel). (a) Channel I and (c) Channel II at $k_x = 0$, (e) degenerate Channel I or II at $k_x = 0.5$. (b, d, f) Orbital-projected band structure of bilayer BlueP, circle size scales with orbital contribution. (b) $p_z$ orbital, (d) $p_y$ orbital and (f) $p_z$ orbital. The scattering state in (a), (c) and (e) is marked with arrow in (b), (d) and (f), respectively.

Once the device enters the tunneling regime ($\Delta m \geq 1$), the transmission becomes weakly dependent on further increases in $\Delta m$ but decays exponentially with the bubble width $M$. For a fixed $M$, the transmission curves remain nearly unchanged as $\Delta m$ increases from 1 to 8 (Fig. 1(b,c)), indicating that additional bending of the top layer has little influence after the layers are sufficiently separated. In contrast, increasing $M$ extends the semiconducting segment along the transport direction, leading to an exponential suppression of tunneling transmission, consistent with a finite-barrier picture. Consequently, the switching contrast quantified by TBR is governed primarily by the bubble width rather than its height (Fig. 1(d)). This distinct geometric dependence implies that device performance is robust against variations in vertical deformation, simplifying the structural design and operational control of these homojunctions.

Besides the overall suppression of transmission, the bubble geometry induces a momentum-selective filtering effect in zigzag homojunctions. For the $Z_{10}^{10}$ device at the Fermi energy, the transverse-momentum-resolved transmission $T(k_x)$ (Fig. 1(c)) reveals that the bubble-free device exhibits quantized transmission of $T = 2$ within two momentum windows ( $k_x \in [-0.18, 0.18]$ and $k_x \in [0.36, 0.64]$ ), corresponding to carriers near the Brillouin-zone center ($k_x = 0$) and near zone edge ($k_x = 0.5$), respectively. Upon introducing a bubble, the transmission near $k_x = 0$ is reduced by a factor of $\sim 20$, whereas the suppression around $k_x = 0.5$ is much stronger, reaching a factor of $\sim 200$. This order-of-magnitude contrast demonstrates that the bubble-defined junction acts as an intrinsic momentum filter that preferentially transmits $\Gamma$-centered carriers.

The microscopic origin of this momentum selectivity is revealed by the scattering states in the bubble region (Fig. 2(a,c,e)). We characterize the attenuation using the decay length $\xi$, defined as the distance over which the wavefunction amplitude decreases to $1/e$ of its initial value. At $k_x = 0$, the two transport channels respond very differently to bubble-induced layer separation. Channel I exhibits substantial amplitude in the interlayer region and becomes strongly attenuated once the layers are separated, leading to a short $\xi = 10.18$ Å. In contrast, Channel II remains largely confined within the individual layers and therefore preserves a much longer $\xi = 38.25$ Å, sustaining relatively high transmission near $k_x = 0$. Around $k_x = 0.5$, the scattering states show comparably short decay lengths ($\xi = 24.59$ Å, Fig. 2(e)), meaning that no similarly robust channel persists, consistent with the much stronger suppression of $T(k_x)$ in that momentum window.

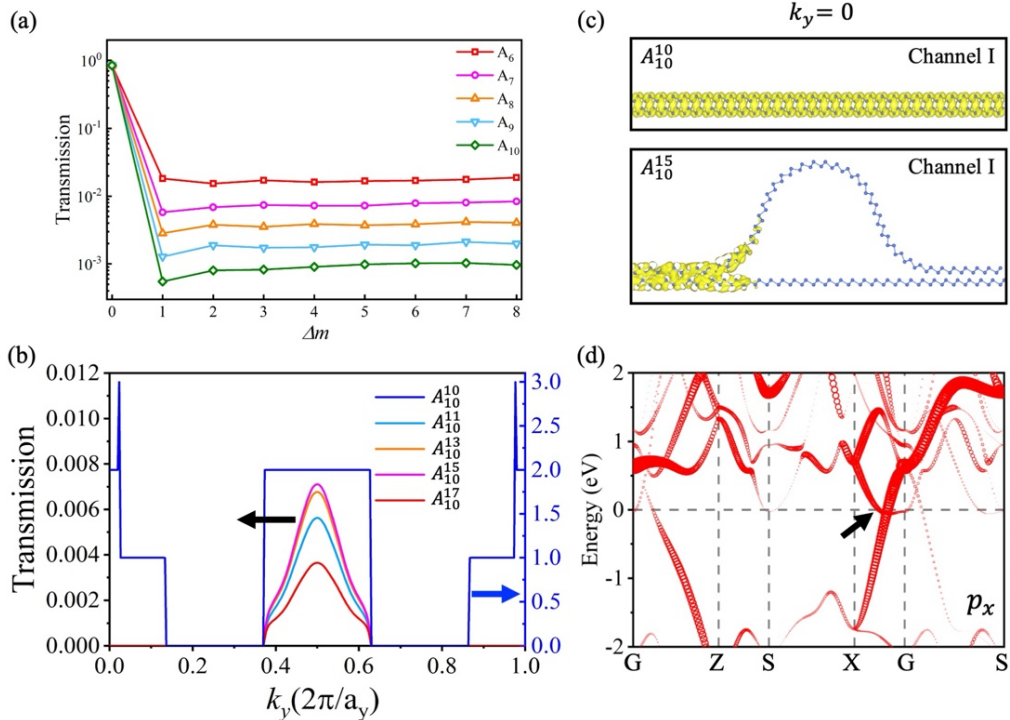

**FIG. 3.** Transport properties of armchair homojunctions. (a) Transmission versus $\Delta m$; (b) Transmission versus $k_y$ for $A_{10}^N$ ($N \in [10,17]$), the blue line corresponds to the right Y-axis, the others to the left (arrows). (c) Scattering state in Channel I at $k_y = 0$; (d) $p_x$ orbital-projected band structure, circle size scalses with orbital contribution. The scattering state in (c) is marked with arrow in (d).

Orbital-resolved analysis provides a direct explanation for the distinct decay behaviors (Fig. 2(b,d,f)). At $k_x = 0$, Channel I originates from interlayer hybridization of $p_z$ orbitals, which produces significant electron density in the interlayer region and is therefore highly sensitive to the bubble-induced separation. By contrast, Channel II at $k_x = 0$ is dominated by intralayer $p_y$ hybridization, yielding charge density localized within each layer and remaining robust against vertical separation, especially in the straight bottom layer. The states contributing near $k_x = 0.5$ mainly arise from intralayer $p_z$ hybridization; while intralayer in nature, they are less robust than the $p_y$-dominated channel. Consequently, they suffer stronger attenuation, leading to the pronounced transmission blocking near $k_x = 0.5$. Taken together, these results establish a clear microscopic transport hierarchy in bubble-deformed bilayers: intralayer-dominated channels are more resilient than interlayer-hybridized channels, and σ-type intralayer bonding (e.g., $p_y$) is more robust than π-type intralayer bonding (e.g., $p_z$) against bubble-induced structural perturbations.

We next investigate the transport properties of armchair-oriented homojunctions. As in the zigzag configuration, these devices undergo a crossover from ballistic to tunneling transport upon bubble formation. As shown in Fig. 3(a), the transmission coefficient at the Fermi level decreases from ~ 0.82 to below $10^{-2}$ when a bubble is introduced. In the tunneling regime, the transmission decays exponentially with the bubble width and exhibits only a weak dependence on the bubble height. Notably, the resistance–width dependence is orientation-dependent. Fitting the width-dependent resistance with an exponential form, $R(W) = e^{aW+b}$, yields $a_A = 0.12, b_A = 1.84$ for armchair junctions, compared with $a_Z = 0.09, b_Z = 2.02$ for the zigzag direction (Fig. S7). The larger fitted $a$ in the armchair orientation implies a systematically higher resistance over the investigated width range. In addition, the armchair junctions display momentum-selective transport (Fig. 3(b)): electron states near $k_y = 0$ are strongly suppressed, whereas those near $k_y = 0.5$ remain largely transmitted. This momentum selectivity originates from the orbital hybridization character of the scattering states. The orbital-projected analysis in Fig. 3(c,d) indicates that the contribution near $k_y = 0$ arises primarily from interlayer-hybridized $p_x$ states (with π-like character), which are readily disrupted by the bubble-induced increase in interlayer separation. By contrast, the states near $k_y = 0.5$ are dominated by intralayer π-type hybridization and are therefore less sensitive to vertical decoupling (Fig. S4). Consequently, the bubble deformation preferentially filters interlayer-coupled modes while preserving intralayer-dominated transport channels.

Building on the geometry-dependent scaling and the orbital-resolved origin of the momentum selectivity discussed above, a practical design criterion for geometry-controlled homojunctions is formulated (Fig. 4). Transport robustness is governed primarily by the spatial origin of hybridization: channels sustained by intralayer coupling are generally more stable than those relying on interlayer coupling, because bubble deformation directly weakens interlayer interactions. Another factor is the bonding character, with σ-type interactions typically supporting more efficient transport than π-type interactions. This hierarchy implies that crystallographic directions dominated by robust intralayer σ bonding are better suited for interconnect applications requiring stable conduction, whereas junctions whose current is mediated by interlayer π-coupling offer stronger geometric tunability and are therefore attractive for switching applications.

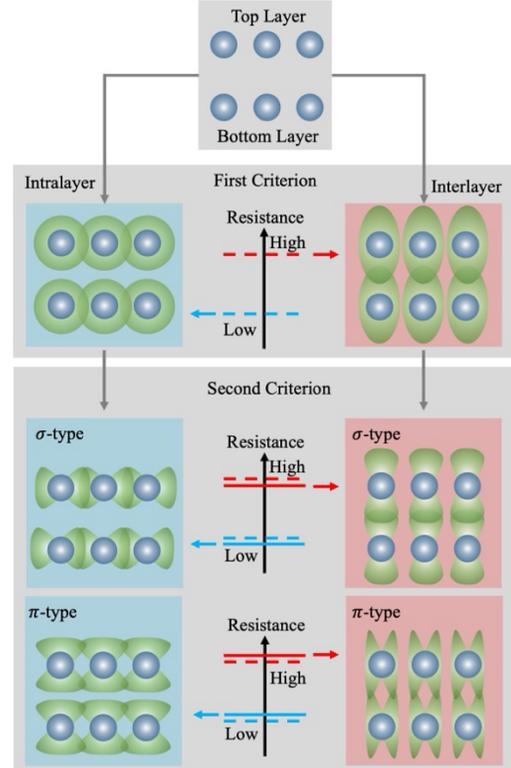

**FIG. 4.** Schematic for evaluating transport robustness.

The mechanism and methodology developed above can be extended to BlueP homojunctions in which bubble-like deformations are introduced in both layers. To examine this scenario, we constructed dual-layer junctions with the two layers bulging in opposite directions (denoted by "*", Figs. S8(a), S9(a)) and in the same direction (denoted by "#", Figs. S8(b), S9(b)). For zigzag homojunctions, the transmission of $Z_{10}^{10*}$ and $Z_{10}^{10\#}$ decrease to 0.0024 and 0.0026, respectively. The close agreement between these values indicates that the transport behavior is largely insensitive to the relative bulge orientation. Both values are, however, substantially smaller than that of the corresponding single-layer-bubble structures $Z_{10}^{N}(N > 10, T = 0.043)$. This additional suppression arises because the bubble in the bottom layer introduces out-of-plane atomic displacements that disrupt intralayer hybridization, whereas the bottom layer remains straight in $Z_{10}^N$ structures. In armchair homojunctions, $A_{10}^{10*}$ and $A_{10}^{10\#}$ exhibit even stronger suppression, with transmission of $3.53 \times 10^{-6}$ and $7.01 \times 10^{-6}$, respectively. The transmission again shows no significant dependence on the relative bulge orientation, yet it drops much more sharply compared with $A_{10}^N (N > 10, T = 0.0015)$. This substantial reduction is likewise associated with the loss of a straight bottom layer; importantly, the $\pi$-type intralayer hybridization governing transport in the armchair orientation is intrinsically less robust and can be nearly quenched by dual-layer out-of-plane deformation. Overall, the weak sensitivity to the bulge orientation suggests a degree of structural resilience against variations in bubble configuration, which is beneficial for achieving reliable performance in practical nanoelectronic devices.

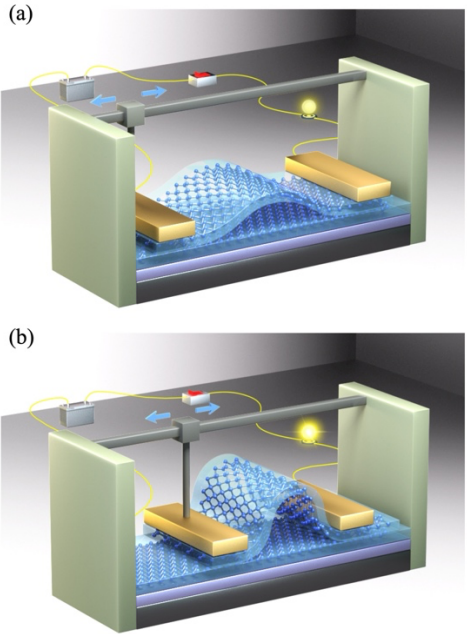

**FIG. 5.** Schematic of a sliding rheostat based on bilayer BlueP: (a) high- and (b) low- resistance states.

Leveraging the pronounced contrast between the two transport regimes identified above, bubble-defined BlueP homojunctions naturally enable electromechanical functionalities. In particular, the bubble-free configuration supports high-conductance, ballistic transport and can serve as the ON state, whereas bubble formation drives the channel into a tunneling-dominated, low-conductance OFF state. This ON/OFF switching modality suggests mechanically switchable memory elements with ON/OFF ratios as high as 30. Furthermore, the exponential dependence of transmission on the effective bubble width, which remains robust across varied structural conditions, further motivates a nanoscale sliding rheostat concept (Fig. 5). In this design, the bottom layer is immobilized on a substrate, while the top layer is allowed to slide laterally and is electrically contacted at both ends by external electrodes. The sliding distance $L$, defined as the separation between the two electrodes, continuously tunes the effective tunneling length and therefore modulates the device resistance exponentially. Specifically, the resistance follows $R_Z = \exp(0.09L + 2.02)$ K$\Omega$ and $R_A = \exp(0.09L + 1.84)$ K$\Omega$ for zigzag and armchair homojunctions, respectively, where $L$ is taken as its numerical value in Ångströms. As established earlier, this resistance–distance dependence is stable and reproducible. The proposed sliding rheostat could therefore function both as a continuously tunable resistive element in nanocircuits and as a sensitive electromechanical sensor capable of resolving Ångström-scale displacements via electrical readout.

In summary, we systematically investigated the electronic transport of BlueP homojunctions incorporating nanoscale bubble deformations. Bubble formation induces a clear crossover from ballistic transport to a tunneling-dominated regime, resulting in a modulation of conductance with an ON/OFF ratio as high as 30. This transition is accompanied by a robust momentum-selective filtering effect in the tunneling regime. In zigzag-oriented junctions, carriers near $k_x = 0$ are preferentially transmitted, while those near $k_x = 0.5$ are strongly suppressed. In armchair-oriented junctions, states near $k_y = 0.5$ remain largely transmitted, whereas those near $k_y = 0$ are markedly attenuated. These behaviors are governed by the orbital hybridization character of the relevant scattering states. Transport channels dominated by intralayer hybridization are generally more resilient than those relying on interlayer hybridization, and $\sigma$-type bonding supports higher conductivity than $\pi$-type bonding. This hierarchy provides a practical criterion for anticipating transport trends based on orbital character.

We further show that the transmission depends exponentially on the bubble width, while remaining largely insensitive to the bubble height and bulging direction. This yields a stable and geometry-tunable transport response. Based on this relationship, we propose a nanoscale sliding rheostat in which the resistance can be precisely and reproducibly controlled by lateral displacement of the top layer. This concept enables both circuit-level resistance tuning and Ångström-scale displacement sensing through electrical readout.

Looking forward, the physical picture and design principles established here are not limited to BlueP. They should extend to other two-dimensional materials and van der Waals heterostructures in which stacking, strain, and interfacial orbital interactions critically shape layer coupling and confinement. Future experimental efforts may focus on controlled fabrication of such bubble structures through strain engineering or atomic force microscopy indentation, together with real-time electrical characterization of the resulting junctions and of the proposed sliding rheostat. It will also be of interest to explore how external electric, magnetic, or optical fields modulate orbital hybridization and the associated transport anisotropy, which may open routes toward multifunctional and tunable nanoelectronic devices.

We gratefully acknowledge the financial support from the Shenzhen Science and Technology Innovation Commission under the Outstanding Youth Project (Grant No. RCYX20231211090126026), the National Natural Science Foundation of China (Grants No. 12474173, 52461160327), Department of Science and Technology of Guangdong Province (Grants No. 2021QN02L820), and the China Postdoctoral Science Foundation (Grants No. 2022M722191). We also acknowledge HZWTECH for providing computation facilities.

## AUTHOR DECLARATIONS
### Conflict of Interest
The Authors declare that they have no known competing financial interests or personal relationships that could have appeared to influence the work reported in this paper.

## DATA AVAILABILITY
The data that support the findings of this study are available from the corresponding authors upon reasonable request.